\documentclass{article}
\usepackage{hyperref}
\hypersetup{
  colorlinks=true,
  allcolors=[RGB]{34,139,34}
}
\usepackage{mdframed}

\begin{document}
\date{}

\title{Note on fair coin toss via Bitcoin}
\author{
Adam Back\\
\small{\texttt{adam@cypherspace.org}}
\and
Iddo Bentov\\
\small{\texttt{idddo@cs.technion.ac.il}}
}

\maketitle

\section{Introduction}

In this short note we show that the Bitcoin network can allow remote parties to gamble with their bitcoins by tossing a fair or biased coin, with no need for a trusted party, and without the possibility of extortion by dishonest parties who try to abort. The superfluousness of having a trusted party implies that there is no house edge, as is the case with centralized services that are supposed to generate a profit.

One simple way to accomplish a coin toss protocol with Bitcoin is via a protocol fork that adds to the Bitcoin scripting language an opcode that puts on the stack the hash of the block in which the transaction resides. However, this implies that the parties have to wait for 10 minutes on average until the result of the bet becomes known. Worse still, such an opcode should have a maturity time of e.g. 100 blocks due to possible reorgs, thus the winning party will have to wait for more than 16 hours before being able to spend the coins that she won.

We propose an alternative coin toss protocol that utilizes the current Bitcoin implementation, i.e. with no need for a protocol fork. Further, with our protocol it is not necessary to wait for the next solved block, and instead the amount of coins of the bet can dictate the appropriate confidence level that the parties require. This means that 0-confirmations security for low value bets does not use the PoW irreversibility property, and instead the mining race degrades into a network race. Hence this is similar to Point of Sale for low value transactions with Bitcoin, as merchants can take a small risk by accepting unconfirmed transactions, while listening on the network to detect double-spending attempts.

\section{Protocol}
The reason why we can resist malicious adversaries who abort upon discovering that they lost the bet is that the Bitcoin scripting language allows us to have a primitive with which Alice locks a certain amount of her coins until a specified time in the future, and Bob can spend these coins to an arbitrary address at any time upon meeting certain arbitrary conditions that were specified in advance via a Bitcoin script, otherwise the locked coins are returned to Alice.

This primitive can be implemented in Bitcoin as follows: Alice creates a ``principle'' transaction that takes inputs that she controls, and can be spent according to "(Alice's signature AND Bob's signature) OR (arbitrary conditions)". Alice keeps the ``principle'' transaction private, and creates another ``refund'' transaction that spends the ''principle'' transaction to an output address that she controls, but has locktime set in the future. Alice then signs the ``refund'' transaction, and sends Bob a private message with the the ``refund'' transaction, asking Bob to sign it. Notice that since Bob only sees the hash of the ``principle'' transaction, he can protect himself from being tricked into signing a malevolent transaction that steals his other coins by generating a fresh secret key and asking Alice to create the ``principle'' transaction with the corresponding public address of this key. Hence, Bob sends Alice a private message with his signature for the ``refund'' transaction. Now Alice broadcasts the ``principle'' transaction to the Bitcoin network. If Bob (or anyone) cannot meet the conditions that the ``principle'' transaction specified, Alice will recover her coins after the locktime expires. \cite{g01}

\medskip
Suppose that Alice and Bob wish to do a fair coin toss where each of them inputs X coins and the winner gets the 2X coins. This can be done by selecting the winner according to the least significant bit of two committed secrets, with the following protocol:
\medskip

\clearpage
\begin{small}
\begin{mdframed}[leftmargin=4mm]
\begin{enumerate}
\setlength{\itemsep}{5pt}
\item Alice picks some random secret $A1$ and sends a private message to Bob with the value $A2=\texttt{SHA256}(A1)$.
\item Bob picks some random secret $B1$ and sends a private message to Alice with the value $B2=\texttt{SHA256}(B1)$.
\item Bob creates a ``bet''transaction that takes as input $2X$ of his own coins, and can be spent by: [Alice's signature AND Bob's signature] OR [$\texttt{SHA256}(A)==A2$ AND $\texttt{SHA256}(B)==B2$ AND (($(A\ \texttt{xor}\ B)\ \texttt{mod}\ 2 == 0$ AND Alice's signature) OR ($(A\ \texttt{xor}\ B)\ \texttt{mod}\ 2 == 1$ AND Bob's signature))]
\item Bob asks Alice to sign a ``refund\_bet'' transaction which spends his $2X$ coins to an address that he controls, and has locktime of (say) 20 blocks into the future.
\item Bob broadcasts the ``bet'' transaction to the Bitcoin network.
\item Alice creates a ``reveal'' transaction that takes as input $X$ of her own coins, and can be spent by: [Alice's signature AND Bob's signature] OR [$\texttt{SHA256}(B)==B2$ AND Bob's signature]
\item Alice asks Bob to sign a ``refund\_reveal'' transaction which spends her $X$ coins to an address that she controls, and has locktime of (say) 10 blocks into the future.
\item Alice broadcasts the ``reveal'' transaction to the Bitcoin network (when she is confident enough that the ``bet'' transaction will not be reversed).
\item Bob redeems the ``reveal'' transaction by revealing B1 (when he is confident enough that the "reveal" transaction will not be reversed).
\item Alice redeems the ``bet'' transaction if she won, otherwise she sends $A1$ to Bob so that he could redeem the ``bet'' transaction without waiting for the locktime to expire.
\end{enumerate}
\end{mdframed}
%\caption{The fair coin toss protocol}
\end{small}
%\end{figure}
\medskip

This protocol is sound because the locktime in step (7) is shorter than the locktime of step (4), therefore Bob cannot cheat by broadcasting the transaction that reveals $B1$ in step (9) right before the locktime of step (4) expires.

\bigskip
By using more bits from the two committed secrets, Alice and Bob can bet with a biased coin so that the party with the worse odds wins the larger amount.

\smallskip
Currently \texttt{OP\_MOD} is considered nonstandard in the Bitcoin protocol, but we can for example combine \texttt{OP\_SHA1} and \texttt{OP\_GREATERTHAN} to gain a similar effect.

%\bigskip
\clearpage
\bibliographystyle{plain}

\begin{flushright}
{\tiny revision 7}
\end{flushright}
\end{document}